\documentclass{emulateapj}

\usepackage{lscape}
\usepackage{apjfonts}
\usepackage{graphicx}
\usepackage{natbib}
\newcommand{\hbeta}{H{$\beta$}}
\newcommand{\halpha}{H{$\alpha$}}
\newcommand{\hdelta}{H{$\delta_{{\rm A}}$}}

\newcommand{\OIII}{[O{\sevenrm\,III}]}
\newcommand{\loiii}{$L_{{\rm [O\,\text{\tiny III}]}}$}
\newcommand{\OIIIa}{[O{\sevenrm\,III}]\,$\lambda$4959}
\newcommand{\OIIIb}{[O{\sevenrm\,III}]\,$\lambda$5007}
\newcommand{\OIIIc}{[O{\sevenrm\,III}]\,$\lambda\lambda$4959,5007}
\newcommand{\NII}{[N{\sevenrm\,II}]}
\newcommand{\NIIb}{[N{\sevenrm\,II}]\,$\lambda$6584}
\newcommand{\SII}{[S{\sevenrm\,II}]}

\newcommand{\SIIab}{[S{\sevenrm\,II}]\,$\lambda\lambda$6717,6731}

 \font\sevenrm=cmr7 scaled 1000

\slugcomment{Accepted to ApJ, November 3, 2009}

\begin{document}

\title{Type 2 AGNs with Double-Peaked \OIII\ Lines:
Narrow Line Region Kinematics or Merging Supermassive Black Hole
Pairs?}

\shorttitle{DOUBLE PEAKED NARROW LINES}

\shortauthors{LIU ET AL.}
\author{Xin Liu\altaffilmark{1}, Yue Shen\altaffilmark{1}, Michael A.
Strauss\altaffilmark{1}, and Jenny E. Greene\altaffilmark{1,2}}

\altaffiltext{1}{Department of Astrophysical Sciences, Princeton
University, Peyton Hall -- Ivy Lane, Princeton, NJ 08544}

\altaffiltext{3}{Hubble Fellow, Princeton-Carnegie Fellow}

\begin{abstract}
We present a sample of $167$ type 2 AGNs with double-peaked
\OIIIc\ narrow emission lines, selected from the Seventh Data
Release of the Sloan Digital Sky Survey. The double-peaked
profiles can be well modeled by two velocity components,
blueshifted and redshifted from the systemic velocity. Half of
these objects have a more prominent redshifted component. In cases
where the \hbeta\ emission line is strong, it also shows two
velocity components whose line-of-sight (LOS) velocity offsets are
consistent with those of \OIII . The relative LOS velocity offset
between the two components is typically a few hundred ${\rm
km\,s^{-1}}$, larger by a factor of $\sim 1.5$ than the line full
width at half maximum of each component. The offset correlates
with the host stellar velocity dispersion $\sigma_{\ast}$.  The
host galaxies of this sample show systematically larger
$\sigma_{\ast}$, stellar masses, and concentrations, and older
luminosity-weighted mean stellar ages than a regular type 2 AGN
sample matched in redshift, \OIIIb\ equivalent width and
luminosity; they show no significant difference in radio
properties.  These double-peaked features could be due to
narrow-line region kinematics, or binary black holes.  The
statistical properties do not show strong preference for or
against either scenario, and spatially resolved optical imaging,
spectroscopy, radio or X-ray followup are needed to draw firm
conclusions.
\end{abstract}


\section{Introduction}\label{sec:intro}

Binary\footnote{We refer to both a pair of SMBHs in a Keplerian
orbit and dual SMBHs at large separations where the galactic
potential dominates as {\it binaries} unless otherwise noted.}
supermassive black holes (SMBHs) are possible outcomes of the
hierarchical mergers of galaxies
\citep[e.g.,][]{begelman80,milosavljevic01,yu02}. In one of the
leading hypotheses, major mergers between galaxies are responsible
for triggering nuclear starbursts and quasar activity
\citep[e.g.,][]{hernquist89}. Despite the success of the merger
scenario in explaining much of the observed phenomenology of AGN
statistics \citep[e.g.,][]{kauffmann00,volonteri03,
wyithe03,hopkins08,shen09} and the cores of massive elliptical
galaxies \citep[e.g.,][]{faber97,kormendy09}, direct observational
evidence for binary SMBHs is surprisingly scarce.

The fraction of binary quasars at separations of tens to hundreds
of kpc (halo) scales, is $\lesssim 0.1\%$ at $1\lesssim z\lesssim
5$ \citep[e.g.,][]{hennawi06,myers08,hennawi09}. On kpc (galactic)
scales there are only a handful of unambiguous low-redshift cases
known, in which both active SMBHs are detected in X-rays
\citep[NGC 6240 and Mrk 463,][]{komossa03,bianchi08}, or in the
radio \citep[3C 75,][]{owen85}. On sub-kpc scales, only one case
is known \citep[0402+379,][]{rodriguez06} of a pair of BHs
detected by VLBI with a projected separation of $\sim 7$ pc. There
are several possible reasons that binary SMBHs are so rare
compared to the expectations from the merger scenario: they spend
most of their time at separations far below kpc scales and hence
are difficult to resolve spatially; one or both of the SMBHs are
heavily obscured; both BHs are rarely active at the same time, and
dynamical differences between a single BH and a binary BH are not
readily discernable at cosmological distances.

Recently, \citet{comerford08} conducted a systematic survey of
type 2 AGNs in the DEEP2 galaxy sample at $\bar{z}\sim 0.6$
\citep{davis03}. They argue that 37 of the 107 AGNs in their
sample are inspiralling binary SMBHs, based on significant
velocity offsets between the narrow-line region (NLR) emission
lines \OIIIc\ and the systemic redshift measured from stellar
absorption features. In particular, two of these objects
\citep[one reported by][]{gerke07} show double-peaked \OIII\ lines
with velocity offsets of a few hundred ${\rm km\,s^{-1}}$ and
spatial offsets of several kpc; they interpreted these as binary
SMBHs when both BHs are active. Such double-peaked \OIII\ emission
line features have also been seen in a few optically-selected AGNs
\citep[e.g.,][]{heckman81,zakamska03,zhou04}.  One advantage of
using spectroscopy to find binary SMBHs is that it is not limited
by spatial resolution, as long as the separation is not so small
that the NLRs are no longer distinct.  On the other hand, such
spectral features may be due to biconical outflows or disk
rotation around a single SMBH
\citep[e.g.,][]{axon98,veilleux01,crenshaw09}. Here we focus on
double-peaked {\it narrow line} objects in which the associated
line emitting gas is on scales of $\sim$100 pc to several kpc.
Therefore in the binary SMBH scenario, our double-peaked sample
would include objects whose two SMBHs along with their own NLR gas
are still co-rotating in the galactic potential, well before
forming a pc-scale binary under dynamical friction
\citep[e.g.,][]{milosavljevic01}.  For double-peaked {\it broad
line} objects which may involve outflows or disks on $\la$pc
scales, see, e.g., \citet{eracleous94} and \citet{strateva03}.

\begin{deluxetable*}{crrrcrrrrrrr}
\tabletypesize{\scriptsize} \tablecolumns{12}
\tablewidth{0pc} \tablecaption {The sample\label{table:sample}}
\tablehead{ \colhead{~~~~~~~~~~~~SDSS Designation~~~~~~~~~~~~} &
\colhead{Plate} & \colhead{Fiber} & \colhead{MJD} & \colhead{$z$}
& \colhead{$\sigma_*$} & \colhead{FWHM$_{{\rm [O\,\,III]},1}$} &
\colhead{FWHM$_{{\rm [O\,\,III]},2}$} & \colhead{$V_{{\rm
[O\,\,III]},1}$} & \colhead{$V_{{\rm [O\,\,III]},2}$} &
\colhead{$V_{{\rm H\beta},1}$} & \colhead{$V_{{\rm H\beta},2}$} }
\startdata
 J$000249.07+004504.8$\dotfill &   $388$ & $345$ & $51793$ & $0.0868$ &  $243$ &  $271$ &  $219$ & $-361$ &  $168$ & $-282$ & $ 197$ \\
 J$000911.58-003654.7$\dotfill &   $388$ & $148$ & $51793$ & $0.0733$ &  $203$ &  $231$ &  $220$ & $-198$ &  $134$ & $-196$ & $ 129$ \\
 J$010750.48-005352.9$\dotfill &   $670$ & $ 55$ & $52520$ & $0.5202$ &  $  0$ &  $281$ &  $645$ & $-546$ &  $ 37$ & $-540$ & $-113$ \\
 J$011659.59-102539.1$\dotfill &   $660$ & $213$ & $52177$ & $0.1503$ &  $181$ &  $304$ &  $203$ & $-159$ &  $145$ & $ -81$ & $ 189$ \\
\enddata
\tablecomments{The full table is available in the electronic
version of the paper. The subscripts ``1'' and ``2'' denote
blueshifted and redshifted components. $V_{{\rm [O\,\,III]}}$ and
$V_{{\rm H\beta}}$ denote velocity offsets relative to the
systemic redshift measured over \OIIIc\ and \hbeta . Velocities
are in units of km s$^{-1}$. The entry reads zero if the quantity
is unmeasurable. The typical uncertainty in $\sigma_*$ is $\sim
15\,{\rm km\,s^{-1}}$. For \OIII, the typical statistical errors
in the best-fit velocity offsets and FWHMs are $\sim 5$ km
s$^{-1}$ and $\sim 10$ km s$^{-1}$. For \hbeta, the statistical
errors are generally a few times larger because \hbeta\ is weaker
than \OIII.}
\end{deluxetable*}

We have carried out a systematic search in the SDSS \citep{york00}
spectroscopic database for emission line AGNs with double-peaked
features or peculiar \OIII\ line shifts relative to the systemic
redshift as measured from stellar absorption features. Here we
report our initial results on a sample of 167 type 2 AGNs with
double-peaked \OIII\ emission lines\footnote{After we submitted
the paper, two similar statistical studies by \cite{smith09} and
\cite{wang09} appeared on arXiv. \cite{smith09} focus on broad
line AGNs with double-peaked \OIII\ lines selected from SDSS
quasar spectra and find a similar fraction ($\sim$1\%) of
double-peaked objects that we found. Because of pipeline
misclassification, their double-peaked sample also contains narrow
line objects, some of which overlap with our catalog.
\cite{wang09} studied a sample of 87 double-peaked narrow line
AGNs of which 42 objects overlap with our sample.}. We describe
our sample selection in \S \ref{sec:data}; the statistical
properties of the sample are presented in \S \ref{sec:study}, and
we discuss our results and conclude in \S \ref{sec:diss}. A
cosmology with $\Omega_m = 0.3$, $\Omega_{\Lambda} = 0.7$, and $h
= 0.7$ is assumed throughout.

\begin{figure}
  \centering
    \includegraphics[width=70mm]{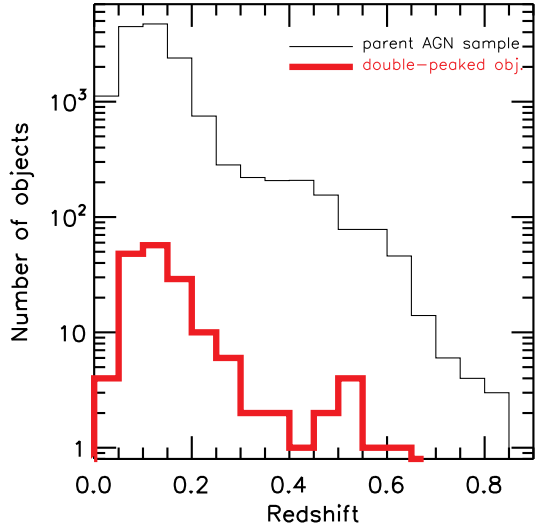}
    \caption{Redshift distribution of the parent AGN and double-peaked samples. }
    \label{fig:redshift}
\end{figure}

\begin{figure*}
  \centering
    \includegraphics[width=50mm]{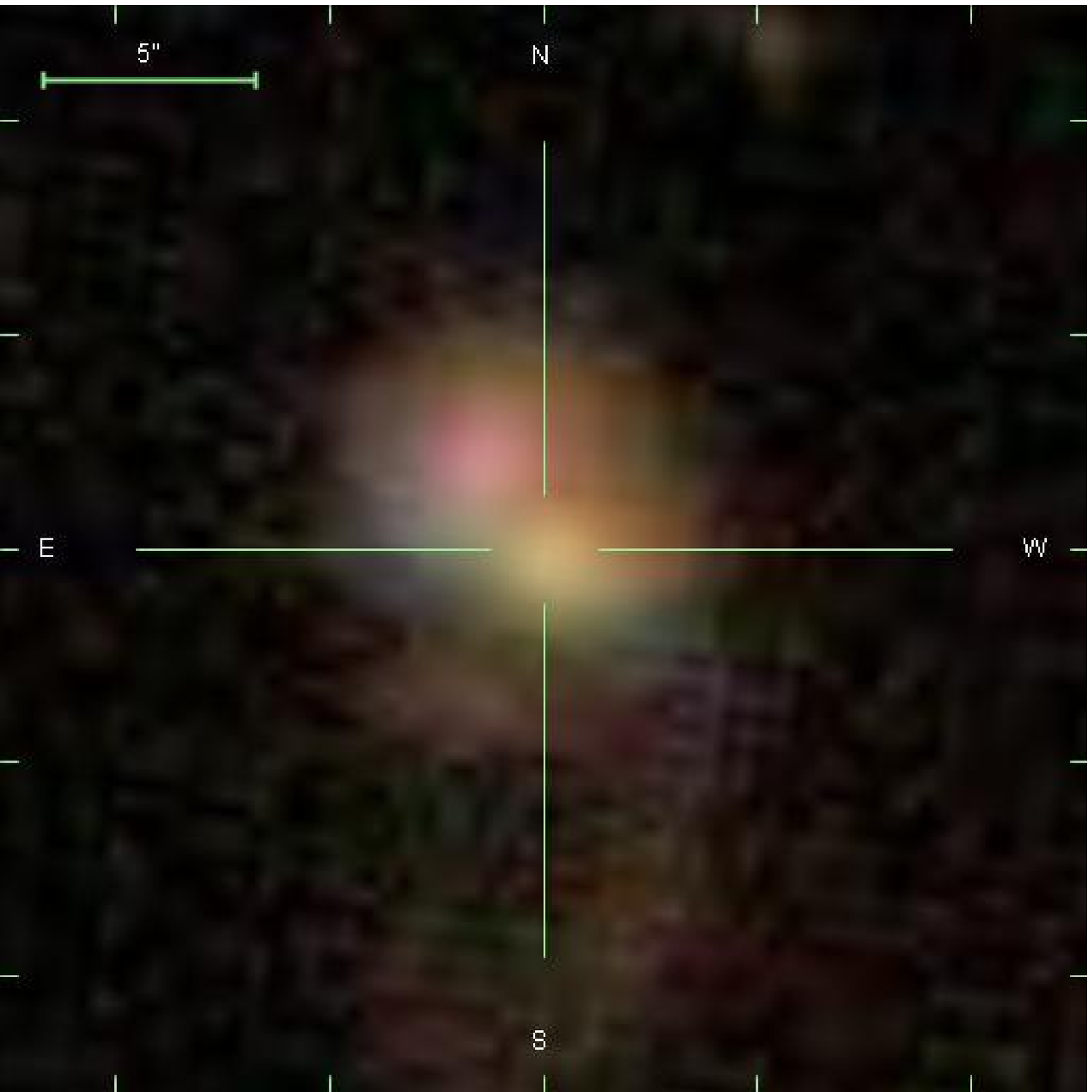}
    \includegraphics[width=90mm]{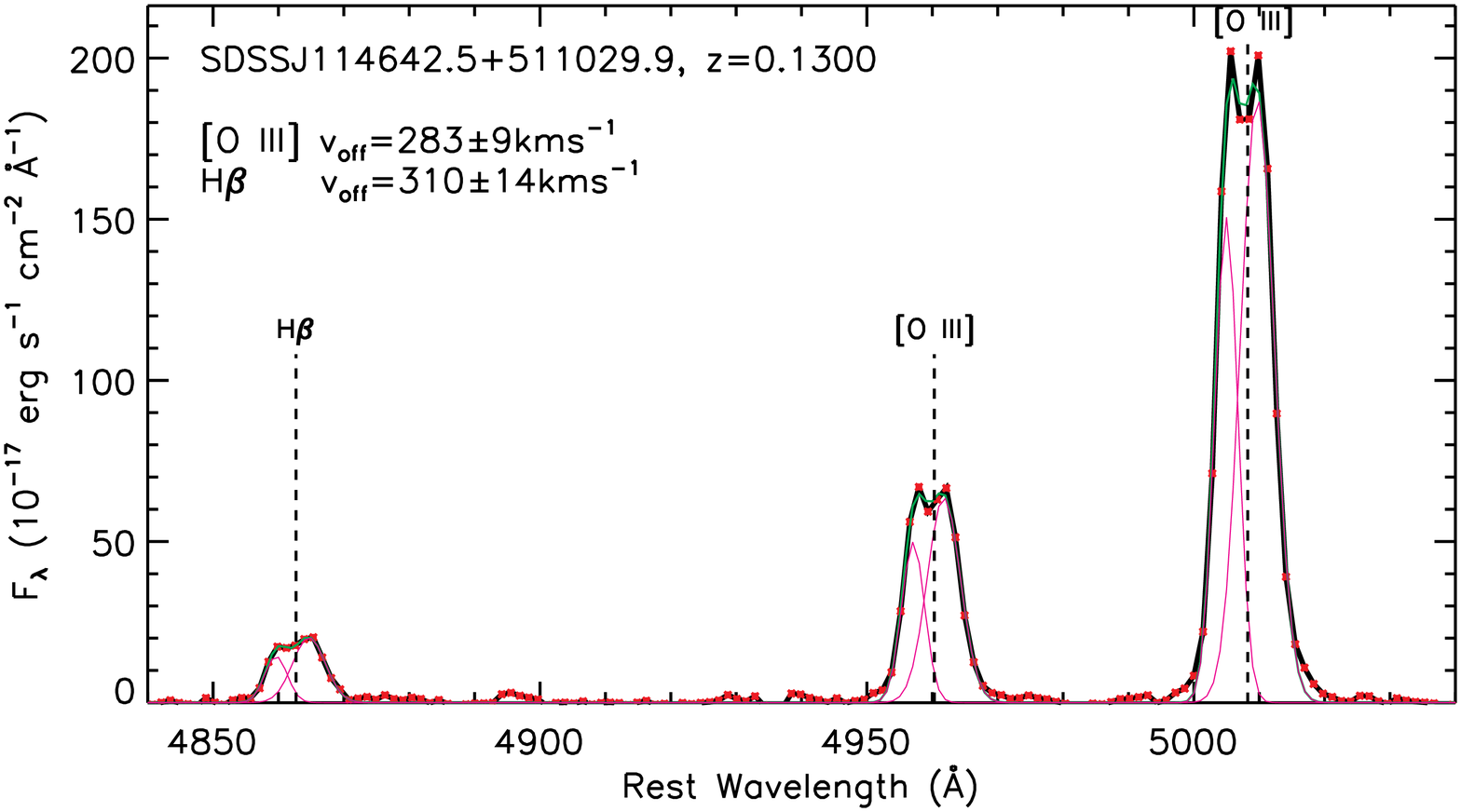}
    \includegraphics[width=50mm]{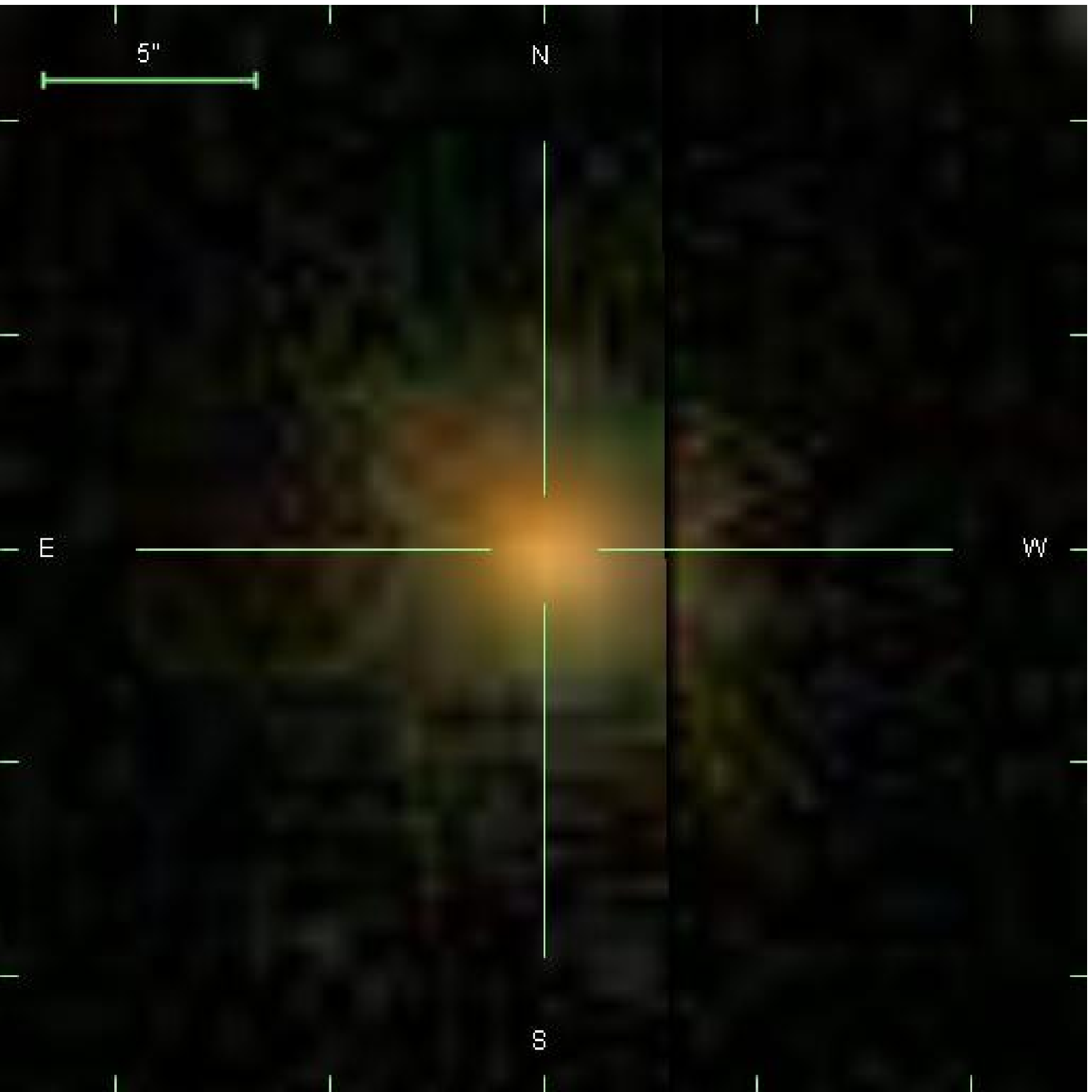}
    \includegraphics[width=90mm]{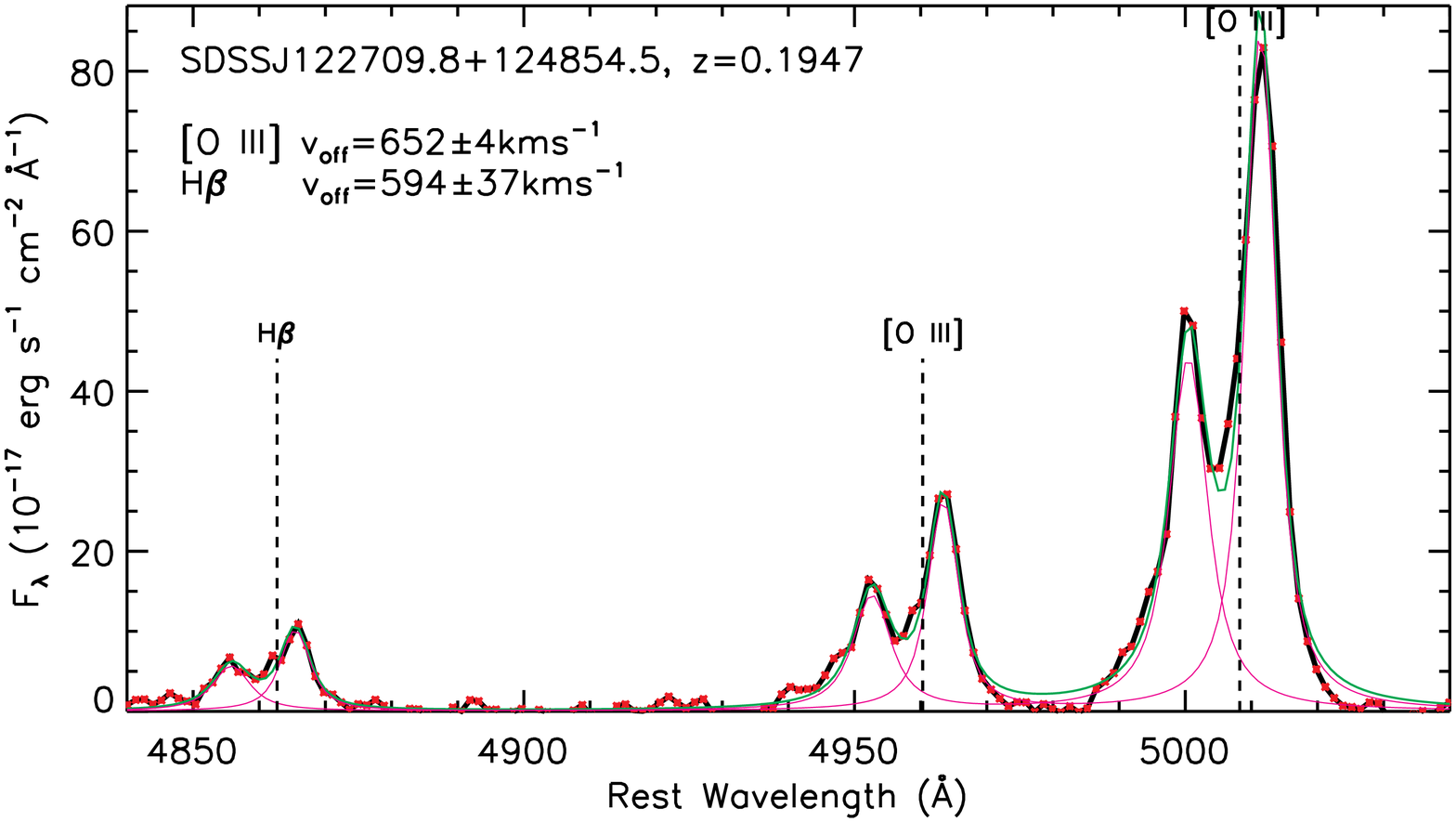}
    \includegraphics[width=50mm]{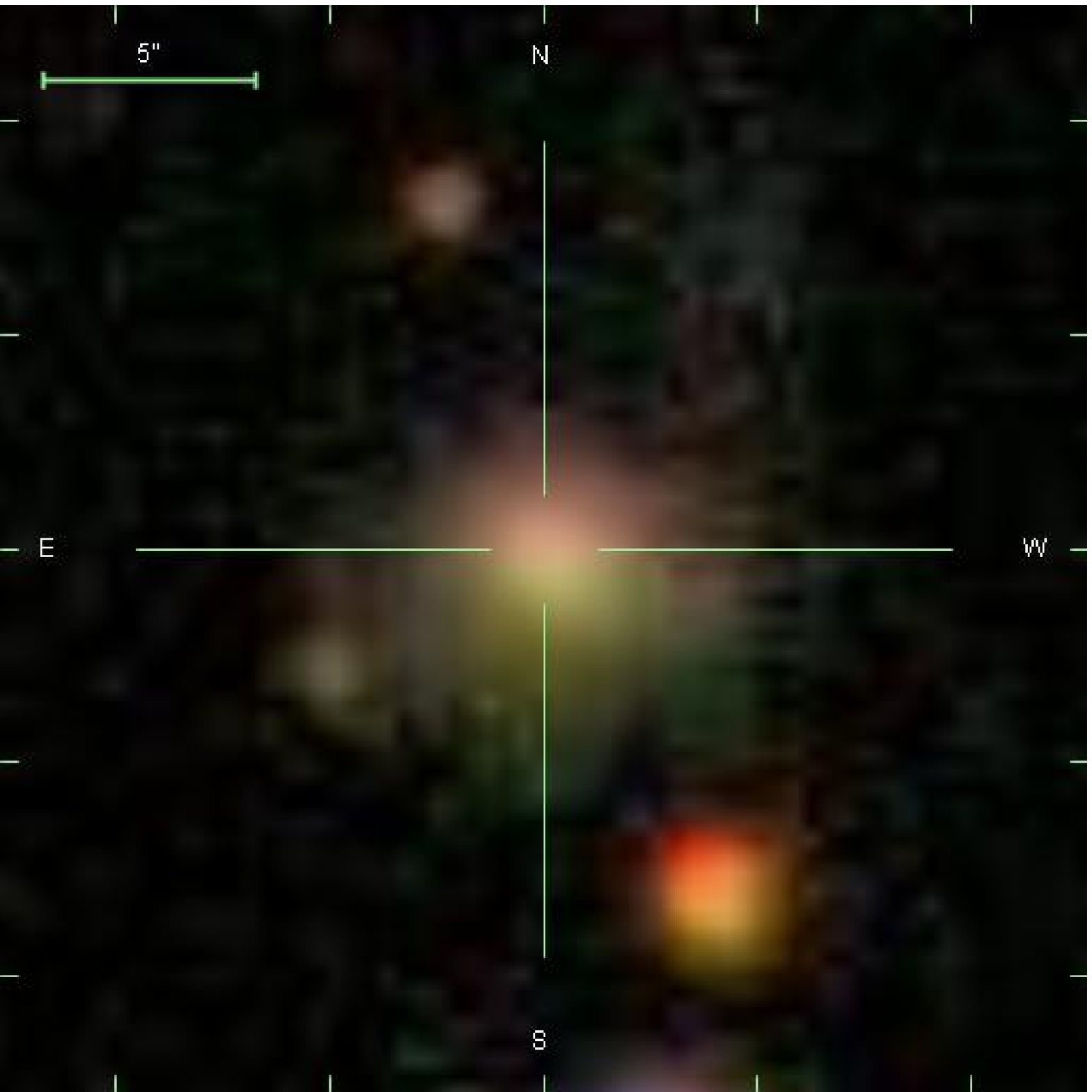}
    \includegraphics[width=90mm]{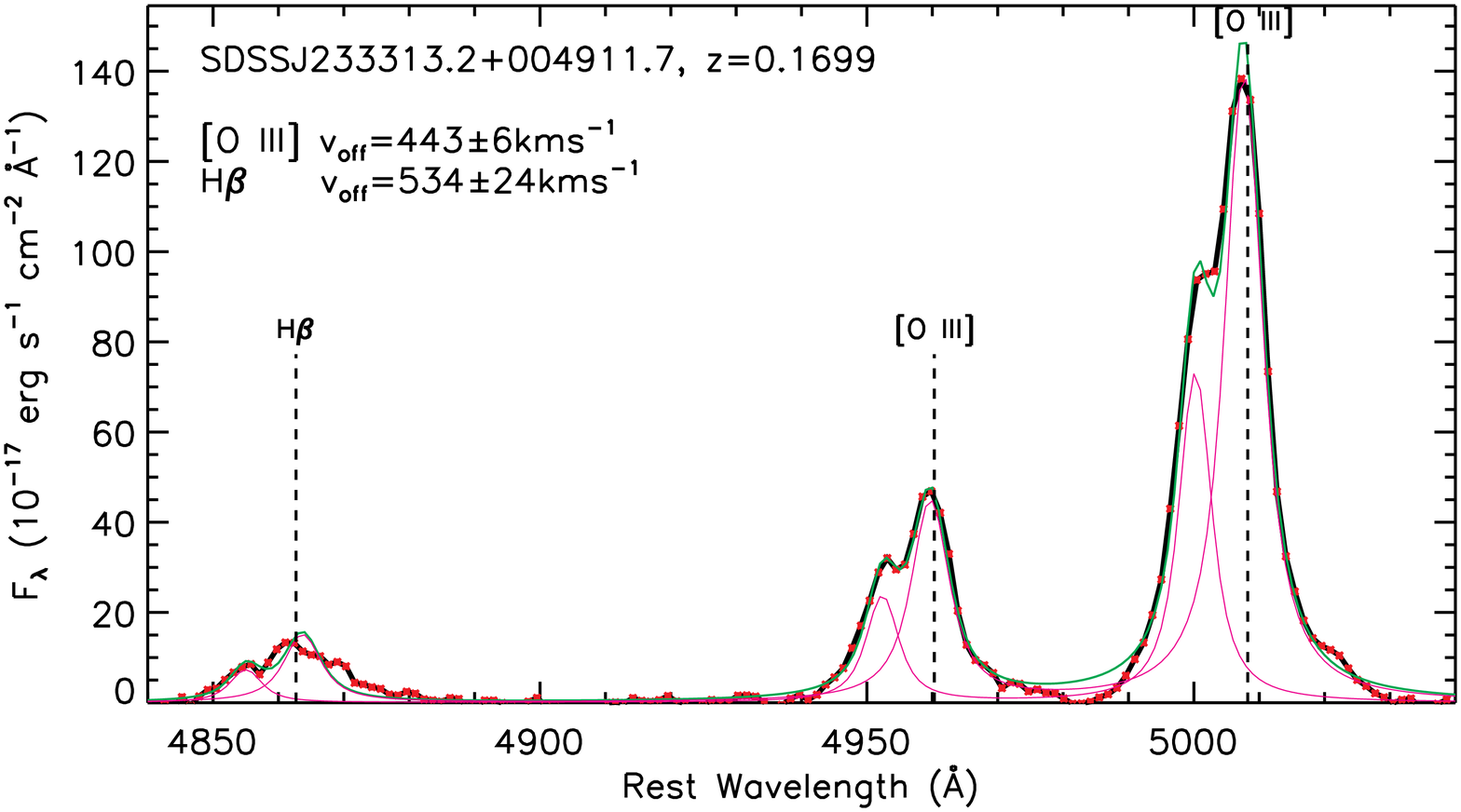}
    \caption{
    Example double-peaked AGNs. Shown here are SDSS $gri$ color-composite images and host-subtracted spectra
    (data points in red and smoothed curve in black) along with
    our best fits (model in green and individual components in magenta) for the
    H$\beta$-\OIII\ region.  The vertical lines are drawn at the
    systemic redshift of each galaxy as measured from stellar absorption features.
    }
    \label{fig:examp}
\end{figure*}

\section{Data}\label{sec:data}

\subsection{The Sample}\label{sec:sample}

Our parent sample is the MPA-JHU SDSS DR7 galaxy
sample\footnote{http://www.mpa-garching.mpg.de/SDSS/} drawn from
the SDSS DR7 spectroscopic database \citep{SDSSDR7}; those include
objects spectrally classified as galaxies by the \texttt{specBS}
pipeline \citep{SDSSDR6} or quasars that are targeted as galaxies
\citep{strauss02,eisenstein01} and have redshifts $z < 0.7$. We
adopt redshifts $z$ and stellar velocity dispersions
$\sigma_{\ast}$ from the \texttt{specBS} pipeline. When
unavailable from the pipeline, we measure $\sigma_{\ast}$ using
the direct-fitting algorithm described in \cite{greene06a} and
\cite{ho09}; we have checked that this code gives results
consistent with those of \texttt{specBS}.  We have checked the
accuracy of the redshift by refitting the stellar continuum with
galaxy templates \citep{liu09}, as well as comparing multiple
epoch spectra of the same objects. The quoted redshift faithfully
traces the stellar absorption features (which we take as the
systemic redshift), with typical statistical errors of a few ${\rm
km\,s^{-1}}$ and systematic errors of $\sim10$ km s$^{-1}$.
Additional spectral and photometric properties such as
emission-line fluxes (determined from equivalent widths and
continuum fluxes), spectral indices, stellar masses, and
concentration indices are taken from the MPA-JHU SDSS DR7 data
product.

We select our AGN sample from this parent sample by the following
criteria: 1) the rest-frame wavelength ranges $[4700,5100]$ \AA\
and $[4982,5035]$ \AA\ centered on the  \OIIIb\ line have median
signal-to-noise ratio (S/N) $>5$ pixel$^{-1}$ and bad pixel
fraction $<30\%$;  2) the \OIIIb\ line is detected at $> 5 \sigma$
and has a rest-frame equivalent width (EW) $>4$ \AA;  3) the line
flux ratio \OIIIb/\hbeta\ $> 3$ if $z > 0.33$, or the diagnostic
line ratios \OIIIb/\hbeta\ and \NIIb/\halpha\ lie above the
theoretical upper limits for star-formation excitation from
\citet{kewley01} on the BPT diagram \citep{BPT} if $z < 0.33$.
This procedure yields $\sim$ 14,300 type 2 AGNs.  We supplement
this sample with $\sim 400$ type 2 quasars from \citet{reyes08}
which are not included in the MPA-JHU data products. These
additional type 2 quasars extend to somewhat higher redshifts ($z
< 0.83$).  Our final AGN sample includes 14,756 objects with high
S/N and high spectral quality around the \OIII\ lines, suitable
for the analysis that follows.  The redshift distribution of this
parent AGN sample is shown in Figure \ref{fig:redshift}.

In many AGNs and starbursts, the narrow forbidden emission lines
\OIIIc\ are known to have peculiarities such as extended blue
wings, velocity offsets from systemic redshifts, and complex line
profiles
\citep[e.g.,][]{heckman81,whittle85,zakamska03,zhou06,komossa08}.
We here focus on a more dramatic subset -- those with
double-peaked \OIII\ emission lines. A complete analysis of the
nature and statistical properties of other peculiar \OIII\ line
characteristics will be presented elsewhere.

The identification of such double-peaked objects requires that
both \OIIIa\ and \OIIIb\ are better fit with two components rather
than a single component.  As the initial screening process we
visually inspected the spectra of all the 14,756 AGNs and
identified 167 objects with unambiguous double-peaked \OIII\
lines. We only include objects which have well-detected double
peaks in both \OIIIa\ and \OIIIb\ with similar profiles; we do not
include those with complex line profiles such as lumpy, winged, or
multi-component features.  We plot the redshift distribution of
the double-peaked sample in Figure \ref{fig:redshift}, and the
SDSS images and spectra of three such examples in Figure
\ref{fig:examp}.  We list all the 167 double-peaked AGNs and their
line measurements in Table \ref{table:sample}.   We caution that
there could be some double-peaked narrow-line objects in SDSS DR7
missed by our selection if the pipeline got the redshifts wrong
because of the peculiar line profiles, although this fraction is
likely to be no larger than a few percent.

\begin{figure*}
  \centering
    \includegraphics[width=170mm]{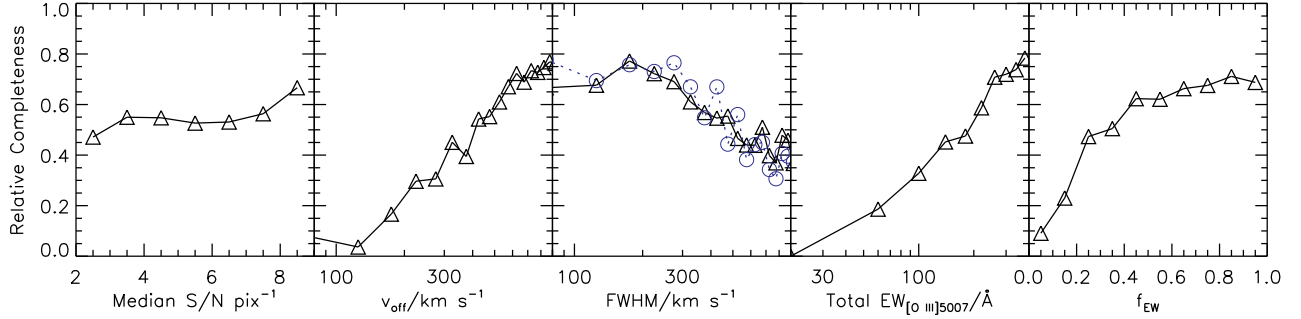}
    \caption{Selection completeness as a function of median $S/N$, $v_{\rm off}$, FWHM, total
    \OIIIb\ EW and the flux ratio between the two components, as determined from our Monte Carlo
    simulations. In the middle panel, black and blue lines are for the two components respectively.}
    \label{fig:comp}
\end{figure*}

We measured the line properties of these double-peaked objects as
follows. The galaxy continuum was subtracted using the best-fit
template constructed from a linear combination of instantaneous
starburst models of \citet{bc03} with ten different ages, as
described in \citet{liu09}.  In the 11 objects for which the
stellar continuum is too weak for measuring $\sigma_{\ast}$ and
for template fitting, a simple power-law model was used.  The
continuum-subtracted \OIII\ region ($\lambda\lambda 4930-5040$\AA)
was fit by a pair of Lorentzian functions convolved with the
measured instrumental resolution of the spectra ($\sigma\sim 65$
km s$^{-1}$).  The redshift and line width for each velocity
component of \OIIIa\ and \OIIIb\ were forced to be the same.  The
flux ratio of \OIIIb\ to \OIIIa\ was allowed to vary (but we found
it was always close to 3).  In cases where \hbeta\ is measurable,
we further fit a double-Lorentzian to the \hbeta\ region
($\lambda\lambda 4850-4880$ \AA), where the positions of the
centroids were allowed to vary, but the widths of the two \hbeta\
components were fixed to the best-fit values for the two \OIII\
components.  We then repeated this fitting process with a
double-Gaussian model, and took as the best fit the model which
had the smaller reduced $\chi^2$ for each object.  We visually
inspected all the fits and verified that the line profiles are
generally well reproduced by the model.
For \OIII, the typical statistical errors in the best-fit velocity
difference between the two components $v_{\rm off}$, line fluxes
and FWHMs are $\sim 5$ km s$^{-1}$, $\sim 5$\%, and $\sim 10$ km
s$^{-1}$, but we caution that the actual uncertainties are likely
to be larger.  For \hbeta, the statistical errors are generally a
few times larger because \hbeta\ is weaker than \OIII.

\subsection{Selection Completeness}\label{sec:sel_comp}

\begin{figure*}
  \centering
    \includegraphics[width=170mm]{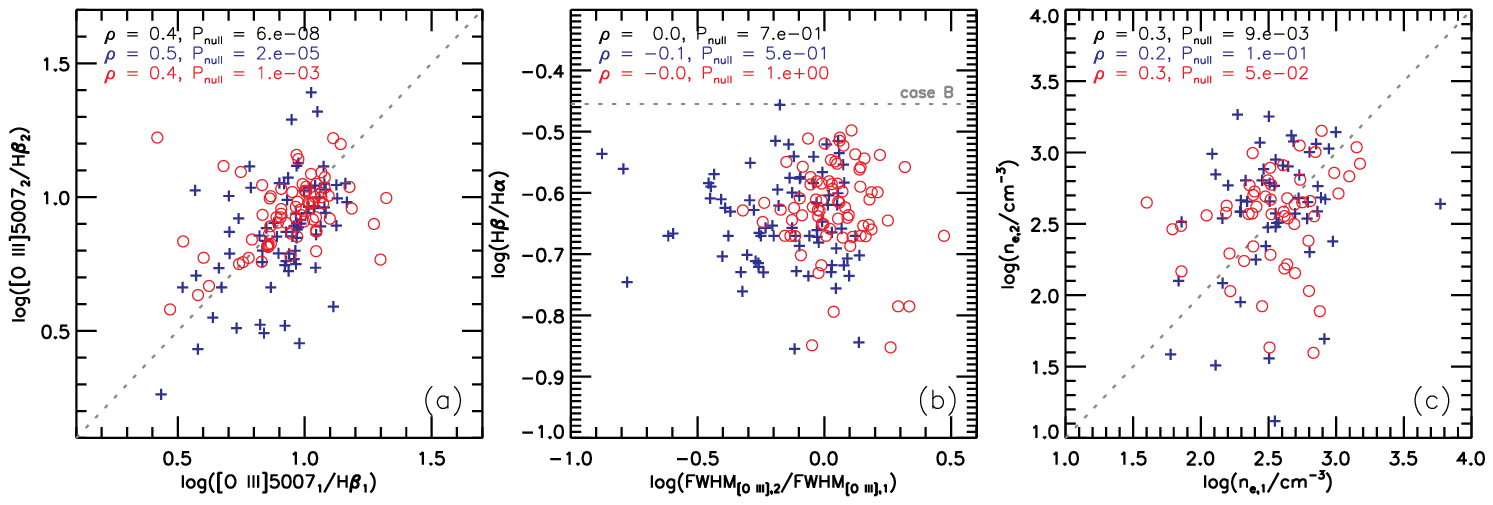}
    \caption{Narrow-line region physical properties of the double-peaked
    components. Objects with stronger redshifted components are plotted as red open
    circles whereas those with stronger blueshifted components are
    plotted as blue plus signs.  The Spearman
    correlation coefficient and the probability of null
    correlation are labelled in each plot for the whole ({\it
    black}), the stronger-blueshifted ({\it blue}), and
    the stronger-redshifted ({\it red}) samples, respectively. (a).
    Ionization parameter (indicated by the flux ratio between \OIIIb\ and \hbeta)
    of the blueshifted component versus that of the redshifted component. (b).
    Line width ratio of the double-peaked components versus Balmer decrement
    (as a measure of reddening). (c). Electron density (inferred from the flux
    ratio between the \SIIab\ doublet) of the blueshifted component versus
    that of the redshifted component.}
    \label{fig:phy}
\end{figure*}

Our selection of double-peaked objects was done by eye and is by
no means objective.  To gain some sense of our completeness we
performed Monte Carlo simulations. We generated mock spectra of
double-peaked objects in the \hbeta-\OIII\ region with random
distributions of continuum S/N, EW, velocity offset, and FWHM of
the two components of \OIII . We randomly assigned a Gaussian or
Lorentzian profile to each line and broadened the spectrum using
an instrumental Gaussian broadening of $\sigma=65$ km s$^{-1}$. In
this way we generated 2,000 mock spectra of double-peaked objects
and mixed them with 18,000 mock spectra of single-peaked objects
with random spectral properties. We then visually inspected the
20,000 mock spectra and identified double-peaked objects in the
same way as we did for the real sample.  We show in Figure
\ref{fig:comp} the selection completeness as a function of median
S/N, $v_{\rm off}$, FWHM, total \OIIIb\ EW, and flux ({\rm EW})
ratio between the two components $f_{\rm EW}$.  In general the
completeness increases with increasing $v_{\rm off}$, \OIIIb\ EW,
and $f_{\rm EW}$, and decreases with increasing FWHM, as expected
from the way our double-peaked objects are identified. As a result
there is a selection bias against objects with small $v_{\rm off}$
and large FWHM that we discuss later. In addition we are very
incomplete for objects with $v_{\rm off}\lesssim 200\, {\rm km\,
s^{-1}}$ partly due to the spectral resolution limit\footnote{For
instance, the object SDSS J100043.13+020637.2, with $v_{\rm
off}\sim 150\,{\rm km\,s^{-1}}$ reported in \citet{comerford09} is
not included in our sample based on the SDSS spectrum alone.}.  On
the other hand, we find that the false detection rate is very low
($< 2\%$ of the double-peaked sample) -- probably an advantage of
visual inspection over automated algorithms.  We note that the
completeness estimated this way is not the {\em absolute}
completeness, which depends on the actual underlying distributions
of all relevant properties. Nevertheless, these completeness
estimates are useful to correct for selection biases. For example,
when other properties are fixed, double-peaked objects with larger
EWs are easier to identify, by a relative amount that can be
determined from our simulation results.

\begin{figure*}
  \centering
    \includegraphics[width=170mm]{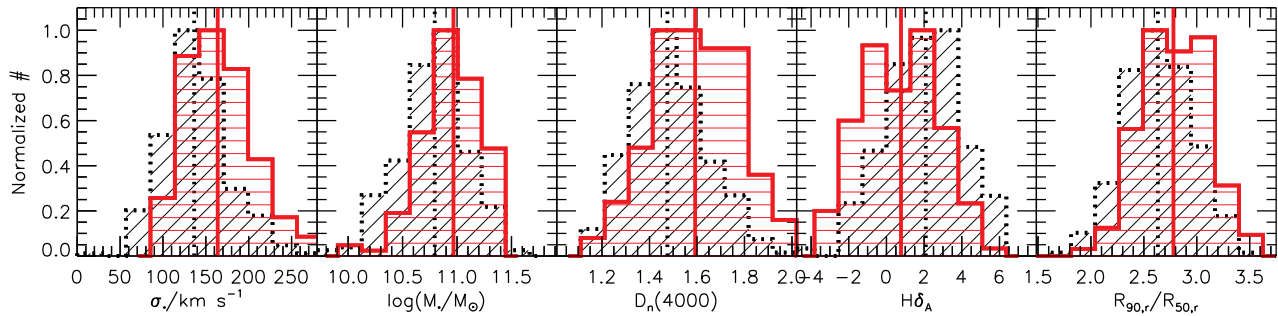}
    \caption{Host galaxy properties (stellar velocity dispersion $\sigma_{\ast}$,
    stellar mass $M_{\ast}$, spectral indices $D_{n}(4000)$ and \hdelta\ indicating
    the luminosity-weighted mean stellar age and the post-starburst fraction, and
    $r$-band concentration index $R_{90,r}/R_{50,r}$) of the double-peaked sample ({\it red}) compared to
    the control sample ({\it black}). The control sample is randomly drawn from the
    parent AGN sample to have identical redshift and \OIIIb\ EW distributions as
    the double-peaked sample. For each distribution the median is marked with a vertical line.}
    \label{fig:hostprop}
\end{figure*}

After accounting for the dependence of selection completeness on
\OIII\ EW, we find tentative evidence that the fraction of
double-peaked objects in the underlying AGN sample depends on
\OIIIb\ luminosity (\loiii ): it increases by a factor of $\sim 2$
from \loiii\ $\sim10^{7.0}L_{\odot}$ to \loiii\ $\sim
10^{8.5}L_{\odot}$, and has no clear dependence on redshift in the
range probed.  In most cases the two velocity components are
redshifted and blueshifted with respect to the systemic redshift
(except that in a few cases, one component is apparently
coincident with the systemic velocity within the errors), and
about half of them have more prominent redshifted components -- a
subset which may be more unusual since objects with outflows may
tend to have more prominent blueshifted components (see below). In
cases where \hbeta\ is measurable, it also shows a double-peaked
feature with a similar velocity offset as that of \OIII. Hereafter
we use $v_{\rm off}$ measured from \OIII\, in our analysis as it
is more robust than that from \hbeta.  As illustrated in Figure
\ref{fig:phy}(a), both velocity components have excitation
diagnostic line ratios \hbeta/\OIII\ characteristic of
AGNs\footnote{For objects with $z < 0.3$, the \NII\ and \halpha\
lines are visible; they appear double-peaked as well, although the
different components often are blended with one another.}. The
FWHM of each velocity component is typically smaller than $v_{\rm
off}$ by a factor of $\sim 1.5$.

\section{Statistical Properties}\label{sec:study}

\subsection{Host Galaxy Properties}\label{sec:host}

Are there any characteristics of the host galaxies of the
double-peaked objects that are different from ordinary AGNs? The
double-peaked sample is not selected uniformly as we demonstrated
in \S \ref{sec:sel_comp}. To account for selection biases, we
construct a control sample from the underlying AGN sample with
identical redshift, \OIIIb\ EW and luminosity distributions as the
double-peaked sample.  We compare their host-galaxy properties,
including stellar mass and velocity dispersion, SDSS colors,
stellar population parameters $D_{n}(4000)$ and \hdelta\
\citep[indicating the luminosity-weighted mean stellar age and the
post-starburst fraction in the past $\sim$ 0.1--1 Gyr,
respectively; e.g.,][]{kauffmann03}, intrinsic extinction
(estimated by the Balmer decrement), effective radius,
concentration (characterized by, $R_{90,r}/R_{50,r}$, the ratio of
radii that contain 90\% and 50\% of the $r$-band Petrosian flux),
and inclination (estimated by the ratio between the major and
minor isophotal axis).  As shown in Figure \ref{fig:hostprop}, the
double-peaked sample has systematically larger stellar velocity
dispersions and masses, older mean stellar ages (and redder
colors), smaller fractions of post-starburst populations in the
past 0.1--1 Gyr, and higher concentrations than the control
sample.  The difference in the centroids of the two distributions
are roughly half of the width of the distributions.
Kolmogorov-Smirnov (KS) tests show that the probabilities that the
two samples are drawn from the same distribution are $P_{{\rm
KS}}<10^{-3}$.  All other properties of the host galaxies we
examined were indistinguishable with KS probabilities $P_{{\rm
KS}}>10^{-1}$.  We also match the double-peaked and the control
samples with the FIRST \citep{becker95,white97} and ROSAT
\citep{voges99,voges00} surveys and find that they contain very
similar fractions of matches with both surveys ($\sim 35$\% with
FIRST and $\sim 2$\% with ROSAT). The subset that has a more
prominent redshifted component is indistinguishable from the rest
of the double-peaked sample in terms of the properties studied
above.

\subsection{Correlations between Line Properties}\label{sec:line}

\begin{figure*}
  \centering
    \includegraphics[width=170mm]{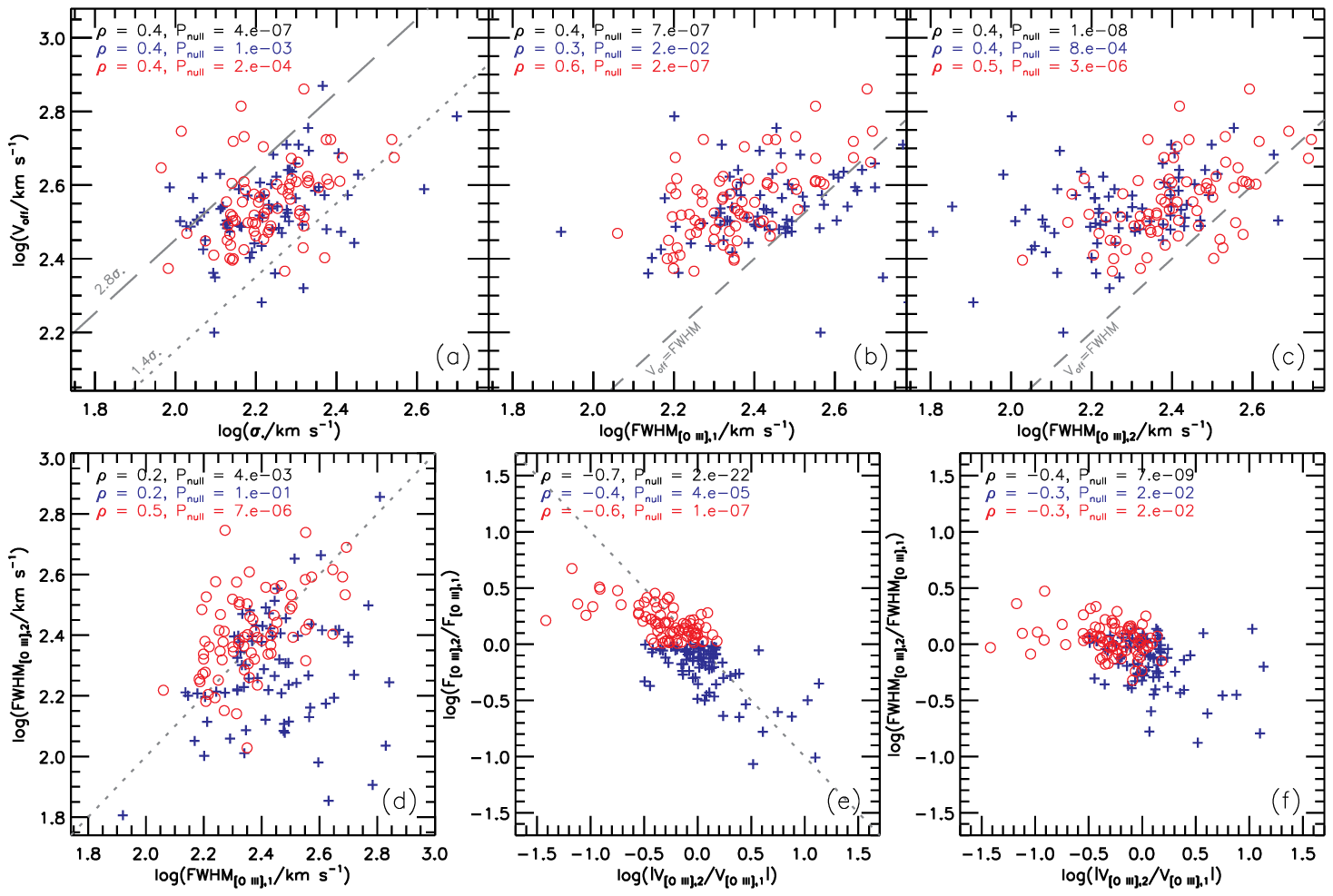}
    \caption{Dynamical properties of the double-peaked components.
    Objects with stronger redshifted components are plotted as red open
    circles whereas those with stronger blueshifted components are
    plotted as blue plus signs.  The Spearman
    correlation coefficient and the probability of null
    correlation are labelled in each plot for the whole ({\it
    black}), the stronger-blueshifted ({\it blue}), and
    the stronger-redshifted ({\it red}) samples, respectively.  (a)--(c).
    stellar velocity dispersion $\sigma_{\ast}$, FWHM of the blueshifted component,
    and FWHM of the redshifted component versus velocity offset between the
    double-peaked components. (d). FWHM of the blueshifted component versus
    that of the redshifted component. (e) Velocity-offset ratio of the
    double-peaked components versus the flux ratio. (f). Velocity-offset
    ratio of the double-peaked components versus the FWHM ratio.}
    \label{fig:line}
\end{figure*}

There are several apparent correlations among the dynamical
properties and the NLR physical conditions of the double-peaked
sample. Figures \ref{fig:phy}(a) and \ref{fig:phy}(c) illustrate
that the two velocity components have similar ionization
parameters (as indicated by the diagnostic line ratio
\hbeta/\OIII) and electron densities (measured from the ratio of
the doublet \SIIab\ for 107 objects in the sample which have low
enough redshifts to have \SII\ coverage in the SDSS spectra and
whose inferred electron densities of both components fall in the
range of [$10$, $10^4$] cm$^{-3}$). Figures \ref{fig:line}(a)--(d)
show that $v_{\rm off}$ is correlated with both $\sigma_{\ast}$
and the line FWHMs of the two components, and the line FWHMs are
correlated with each other. In all these four relations, the
correlation for objects with more prominent redshifted components
appears somewhat stronger than that for the rest. There is some
selection bias at the low $v_{\rm off}$ versus large FWHM corner,
where the double-peaked feature is difficult to identify. However,
our Monte-Carlo simulations show no selection bias against objects
with high $v_{\rm off}$ and small FWHMs (\S\ref{sec:sel_comp}).
In particular, the correlation between $v_{\rm off}$ and
$\sigma_{\ast}$ suggests that the double-peaked line-emitting
regions are on galactic scales where the galactic potential
dominates the bulk kinematics.  Furthermore, Figures
\ref{fig:line}(e) and \ref{fig:line}(f) show that both the \OIII\
line flux ratio and the line width ratio of the two velocity
components are anti-correlated with the ratio of their LOS
velocity offsets relative to the systemic redshift \citep[see
also][]{wang09}, which most likely results from momentum
conservation.

\section{Discussion}\label{sec:diss}

\subsection{NLR Kinematics}

The double-peaked feature may result from particular NLR
geometries such as biconical outflows\footnote{In principle, a NLR
with inflowing gas and dust can also produce the observed
features, but this scenario seems less favored considering the
difficulty of removing high angular momentum given the large
$v_{\rm off}$ \citep[e.g.,][]{heckman81}.} or rotating disks on
kpc scales.  In these scenarios, there is only one SMBH and the
observed blueshifted and redshifted emission-line peaks arise from
NLR gas moving towards and away from us.  Certain nearby Seyfert
galaxies known to have biconical outflows show double-peaked
emission when spectra are taken over the whole galaxy.  Such
prototypical cases include NGC 1068 \citep[e.g.,][]{axon98} and
NGC 3079 \citep[e.g.,][]{duric88}.  It is very likely that there
are some such sources in our sample.  Several objects in our
sample show clear galactic disks in their SDSS images.  Here the
correlation between line width and velocity shift naturally
arises, as the bulge gravitational potential well influences the
motion of NLR gas.  The correlation between the diagnostic line
ratio \OIII /\hbeta\ of the two components shown in Figure
\ref{fig:phy}(a) may be explained if the NLR gas of both velocity
components is ionized by a single SMBH.

Can all the objects in our double-peaked sample be ascribed to NLR
geometry? Outflows and inflows of NLR gas have long been invoked
to explain several commonly observed features in NLR high
ionization lines (such as \OIII), including extended wings
(usually blueshifted from the systemic redshift), line
asymmetries, and broader line widths than those of low ionization
lines \citep[e.g.,][]{heckman81,penston84,rodriguezardila06}. In
the cases of NGC 1068 \citep[e.g.,][]{axon98} and Mrk 78
\citep[e.g.,][]{heckman81,pedlar89,whittle04} known to have
biconical outflows, the velocity splitting relative to the
systemic redshift is larger on the blueshifted component, which is
also brighter and has a larger width than the redshifted
component.  The statistical significance of these trends is
unclear, however \citep[e.g.,][]{whittle88}.  Unlike these, our
double-peaked sample has comparable line widths in the blueshifted
and redshifted components as shown in Figure \ref{fig:line}(d);
half of our double-peaked objects have a more prominent redshifted
component than the blueshifted component. In addition, our sample
has consistent $v_{\rm off}$ in \hbeta\ and in \OIII\ whenever
\hbeta\ is measurable, with no significant ionization
stratification as expected in NLR outflows. These possible
discrepancies may arise because our sample has appreciably larger
\OIII\ luminosity than these previous low redshift comparison
samples, and these properties may be a function of luminosity or
redshift or both; alternatively, they may suggest that not all
objects in our double-peaked sample exhibit NLR outflows.

Further possible clues come from host galaxy and line properties.
The FIRST detected fraction for the double-peaked objects is
comparable to that of the control sample.   The host-galaxy
inclination distribution shows no significant difference either,
whereas our selection of a large LOS $v_{\rm off}$ for the double
components would bias towards either edge-on or face-on
inclinations for the rotating disk or bipolar outflow populations,
unless the outflow orientation is not closely linked to the
angular momentum axis of the host galaxies.   Radio outflows in
nearby Seyfert galaxies and low-ionization AGNs indeed appear to
orient randomly with respect to the host galaxy axes
\citep[e.g.,][]{ulvestad84,gallimore06}.   In addition, while a
correlation between $v_{\rm off}$ and line FWHM is observed in NLR
outflows \citep[e.g.,][]{komossa08}, the ratio of velocity shift
to line width is of order unity, unlike the ratio of $\sim$ 1.5
seen in our double-peaked sample. Finally, as shown in Figure
\ref{fig:phy}(b), there is no obvious correlation in our sample
between the Balmer decrement and the line-width ratio between the
two components of the double-peaked feature (as a proxy for
asymmetry).  Such a correlation might exist if the blueshifted and
redshifted NLR gas components are intrinsically symmetric and the
apparent asymmetry results from extinction. \citet{heckman81}
observed a correlation between the asymmetry of a single line
profile (which, however we caution, is different from our
asymmetry indicator -- the line width ratio of the double-peaked
components) and Balmer decrement in a sample of 36 nearby Seyferts
thought to contain outflows, although later studies on nearby
Seyferts \citep[e.g.,][]{whittle85b} found no strong correlations.

\subsection{Merging SMBH Pairs}

Another explanation for the double-peaked features is merging SMBH
pairs. In this scenario, both BHs, along with their own NLR gas
(with scales of order hundreds of pc), are co-rotating in the
galactic potential on $\gtrsim$ kpc scales, well before dynamical
friction causes their orbit to decay and form a gravitationally
bounded compact binary \citep[e.g.,][]{milosavljevic01}.  The
$v_{\rm off}$-$\sigma_{\ast}$ and $v_{\rm off}$-FWHM correlations
in this scenario would naturally arise from dynamics.  The
comparable numbers of objects with more prominent redshifted and
blueshifted components and the null correlation between the Balmer
decrement and the double-peak asymmetry may both be accommodated
if the two emission components are associated with two SMBHs.  The
similar distribution of host-galaxy inclination angles to the
control sample could be explained, given that there is no
preference for the rotation axis of galaxies in the merging
SMBH-pair scenario.

On the other hand, the correlation between the diagnostic line
ratio \OIII /\hbeta\ of the two components shown in Figure
\ref{fig:phy}(a) may be difficult to explain if the NLR gas of
each component is primarily ionized by its own BH.  In addition,
the higher galaxy concentration for objects with double peaks
(Figure \ref{fig:hostprop}) seems inconsistent with these objects
being in the early stage of mergers, although we may be biased to
more compact objects by the necessity to get both merging objects
within the 3-arcsec fiber aperture.  The older mean stellar ages
and the smaller post-starburst fraction in the past 0.1--1 Gyr
also seem counterintuitive, if there is excess star formation in
mergers.  However, these differences might also result from the
mass-age correlation \citep[e.g.,][]{kauffmann03} given that the
median $\sigma_{\ast}$ of the double-peaked sample is larger than
that of the control sample (Figure \ref{fig:hostprop}), or if the
associated star formation is on a different time scale from what
is probed by the $D_{n}(4000)$ and \hdelta\ indices adopted here,
or it is suppressed by AGN activity.

In view of the above arguments, we conclude that our double-peaked
sample may contain a population of NLR outflows or rotating disks,
as well as objects that are merging SMBHs; The statistical
properties of the sample do not show any strong preference for or
against either scenario.  In Figure \ref{fig:examp} we show
several intriguing cases. The first example
(SDSSJ114642.5$+$511029.9) is a merging system of which the
off-centered galaxy is partially covered by the 3-arcsec diameter
fiber. We have checked that the two galaxies are both detected in
the $K_s$ band by 2MASS \citep{skrutskie06}, and have found
evidence that both are AGNs using spatially resolved longslit
spectroscopy, the results of which will be presented elsewhere.
Only 13 of the 167 objects in our double-peaked sample show
possible evidence for two cores in the SDSS images within the
3-arcsec diameter of the spectroscopic fibers. The second object
(SDSSJ122709.8$+$124854.5) does not have a resolved double core in
its SDSS image, and its emission lines have well separated peaks,
symmetrically shifted from the systemic velocity, and the
redshifted component is more prominent. High resolution HST
imaging may help reveal potential double cores unresolved in SDSS
images \citep[e.g.,][]{comerford09}.  For example, one of our
double-peaked objects (SDSSJ130128.8$-$005804.3, at $z = 0.2455$)
shows no double cores in its SDSS image, but \citet{zakamska06}
used HST imaging to show that it consists of two galaxies with a
projected separation of 1.3 arcsec (corresponding to 5.0 kpc)
which well fits into the SDSS 3-arcsec diameter fiber. We will
need spatially resolved spectroscopy to determine whether both of
the galaxies are AGN. The last object in Figure \ref{fig:examp}
has a strong component coincident with the systemic redshift and a
weaker blueshifted component and an even weaker redshifted
component (not fitted), which could arise from a classic NLR
region plus bipolar outflows. Spatially-resolved optical imaging,
spectroscopy, radio and/or X-ray followup are still needed to help
draw firm conclusions on the nature of these double-peaked narrow
line objects.

\acknowledgments

We thank J. Krolik and W. Voges for helpful comments, and an
anonymous referee for a careful and useful report that improves
the paper.  X.L., Y.S., and M.A.S. acknowledge the support of NSF
grant AST-0707266.   Support for J.E.G. was provided by NASA
through Hubble Fellowship grant HF-01196 awarded by the STSI,
which is operated by the Association of Universities for Research
in Astronomy, Inc., for NASA, under contract NAS 5-26555.

Funding for the SDSS and SDSS-II has been provided by the Alfred
P. Sloan Foundation, the Participating Institutions, the National
Science Foundation, the U.S. Department of Energy, the National
Aeronautics and Space Administration, the Japanese Monbukagakusho,
the Max Planck Society, and the Higher Education Funding Council
for England. The SDSS Web Site is http://www.sdss.org/.

The SDSS is managed by the Astrophysical Research Consortium for
the Participating Institutions. The Participating Institutions are
the American Museum of Natural History, Astrophysical Institute
Potsdam, University of Basel, University of Cambridge, Case
Western Reserve University, University of Chicago, Drexel
University, Fermilab, the Institute for Advanced Study, the Japan
Participation Group, Johns Hopkins University, the Joint Institute
for Nuclear Astrophysics, the Kavli Institute for Particle
Astrophysics and Cosmology, the Korean Scientist Group, the
Chinese Academy of Sciences (LAMOST), Los Alamos National
Laboratory, the Max-Planck-Institute for Astronomy (MPIA), the
Max-Planck-Institute for Astrophysics (MPA), New Mexico State
University, Ohio State University, University of Pittsburgh,
University of Portsmouth, Princeton University, the United States
Naval Observatory, and the University of Washington.

Facilities: Sloan

\bibliography{binaryrefs}


\end{document}